\documentclass[aip,jcp,preprint,citeautoscript,groupedaddress,longbibliography]{revtex4-1}
\usepackage{mathtools}
\usepackage{graphicx}

\newcommand{\kB}{k_{\mathrm{B}}}
\newcommand{\kT}{\kB T}

\newcommand{\vek}[1]{\boldsymbol{#1}}          
\newcommand{\nvek}[1]{\boldsymbol{\hat{#1}}}   
\newcommand{\dif}{\mathrm{d}}                  
\newcommand{\me}{\mathrm{e}}                   
\newcommand{\eexp}[1]{\me^{\displaystyle #1}}
\newcommand{\mean}[1]{\left<#1\right>}
\newcommand{\abs}[1]{\left|#1\right|}
\newcommand{\ffrac}[2]{\frac{\displaystyle #1}{\displaystyle #2}}

\newcommand{\Sflat}{S^{\text{(2d)}}}
\newcommand{\Semb}{S^{\text{(emb)}}}
\newcommand{\Savg}{S^{\text{(3d)}}_{\text{avg}}}
\newcommand{\Ssphere}{S^{\text{(sph)}}}
\newcommand{\Sinv}{S^{\text{(inv)}}}
\newcommand{\qc}{q_\text{c}}

\newcommand{\Rmin}{R_{\text{min}}}

\makeatletter
\newif\ifpreprintoption
\@ifclasswith{revtex4-1}{preprint}{\preprintoptiontrue}{\preprintoptionfalse}
\makeatother
\newcommand{\ifpreprint}[2]{\ifpreprintoption #1\else #2\fi}

\begin{document}


\title{
Relating the structure factors of two-dimensional materials in planar and spherical geometries
}


\author{Yongtian Luo}
\author{Lutz Maibaum}
\email{maibaum@uw.edu}
\affiliation{Department of Chemistry, University of Washington, Seattle WA 98195}


\date{\today}

\begin{abstract}
Scattering structure factors provide essential insight into material properties and are routinely obtained in experiments, computer simulations, and theoretical analyses. Different approaches favor different geometries of the material. In case of lipid bilayers, scattering experiments can be performed on spherical vesicles, while simulations and theory often consider planar membrane patches. We derive an approximate relationship between the structure functions of such a material in planar and spherical geometries. We illustrate the usefulness of this relationship in a case study of a Gaussian material that supports both homogeneous and microemulsion phases. Within its range of applicability, this relationship enables a model-free comparison of structure factors of the same material in different geometries. 
\end{abstract}


\maketitle 


\section{\label{sec:introduction}Introduction}

Many important properties of two-dimensional materials can be probed in scattering experiments that measure the structure factor, which contains information about spatial correlations in the material. Computing such structure factors is therefore a common goal also of computational and theoretical studies. A direct comparison between structure factors obtained by these routes can be difficult if different material geometries are used either due to necessity or convenience. In this work we aim to enable such a comparison of structure factors between planar and spherical materials.

Our interest is motivated by the study of mixed lipid bilayers, which serve as model systems that mimic cellular membranes. Depending on lipid composition and thermodynamic parameters these systems exist in different phases. For example, it is well established that some ternary mixtures of phospholipids and cholesterol undergo a transition upon cooling from a homogeneous fluid at high temperatures to a phase-separated state in which the bilayer separates into coexisting liquid-ordered and liquid-disordered phases~\cite{Veatch03, Baumgart03, Scherfeld03, Veatch05, Marsh09}. Such thermodynamic phase separation creates lateral structure in the membrane on a length scale that increases with the size of the system. Organization on smaller length scales can be generated in structured phases such as microemulsion or modulated phases~\cite{Kumar99, Schick12, Shlomovitz14,Shimobayashi16} as well as in systems that display nanoscopic domains~\cite{Feigenson09,Konyakhina11,Rheinstadter13, Armstrong13,Heberle13b,Toppozini14,Nickels15,Enoki18}. Identifying the nature of such lateral organization and understanding the mechanisms that give rise to them continues to be a challenge in membrane biophysics.

An important observable that characterizes the structure of materials is the density correlation function or its Fourier transform, the scattering structure factor~\cite{HansenMcDonald13}. The latter can be directly measured in scattering experiments using either neutrons or X-rays~\cite{Pabst10, Marquardt15}, which makes the density correlation function a principal objective both of theoretical studies~\cite{Hirose12b,Schick12,Shlomovitz13, Palmieri13} and of molecular mechanics computer simulations~\cite{Rosetti12, Ackerman15, Baoukina17, He18} of multicomponent membranes. The integration of results from experiment, theory, and simulation has significantly  enhanced our understanding of the structure of mixed lipid bilayers.

The comparison of structure factors obtained using these three approaches can be hampered by the use of different bilayer geometries. For example, a series of recent neutron scattering experiments have been performed on small, unilamellar vesicles (SUVs)~\cite{Pencer05b, Heberle13b, Nickels15}. The majority of computer simulations, on the other hand, consider planar, periodically replicated patches of bilayer material. Structural correlation functions obtained in these different geometries cannot be directly compared, even if the change in geometry did not affect intrinsic membrane properties: the difference in shape between the spherical and the planar system itself leads to significant differences in measured or calculated structure factors.

Our goal is to derive a mathematical relationship between the structure factor of a spherical system and that of a planar system. We consider only the immediate effect of system geometry on the scattering intensity, and neglect potential changes of the material properties due to system shape. The latter can play an important role in membrane systems if the radius of a vesicle is small because the distribution of lipids is known to be sensitive to the local curvature and variations therein between the inner and the outer leaflet~\cite{Jiang08, Sakuma11, CallanJones11}. With this caveat in mind we derive in the following section an approximate transformation that converts the structure factor of a spherical system into the in-plane structure factor of a planar system of the same material. The utility and the limitations of this transformation are explored in Section~\ref{sec:gaussianmaterial} where we study as an example a material whose properties are described by an extended Landau-Ginzburg model that supports both unstructured and structured fluid phases. General aspects and potential extensions of our approach are discussed in Section~\ref{sec:discussion}.

\begin{figure}[tb]
\includegraphics[width=\columnwidth]{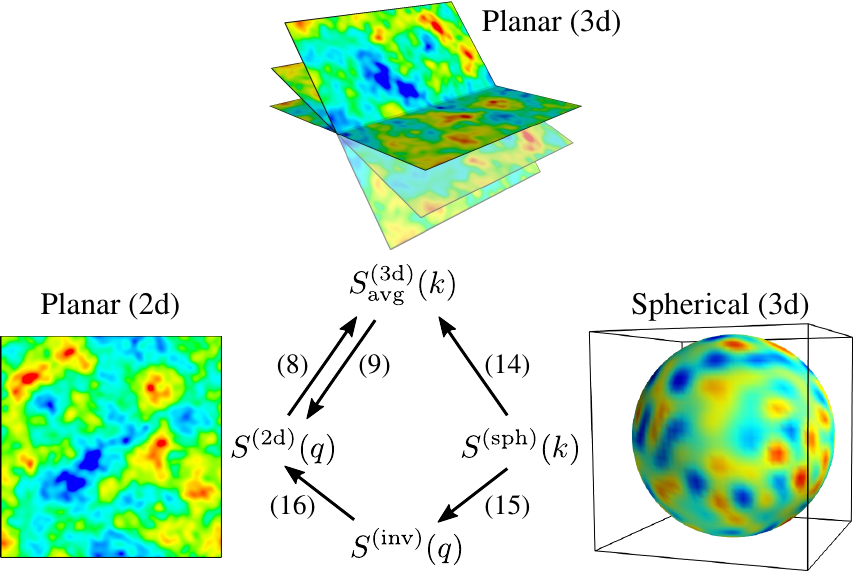}
\caption{\label{fig:sketch}Illustration of the geometries considered in this work. The planar system has two-dimensional structure factor $\Sflat(q)$. Embedding in three-dimensional space and averaging over all possible orientations gives rise to the structure factor $\Savg(k)$. In a spherical geometry this material has three-dimensional structure factor $\Ssphere(k)$, which approaches $\Savg(k)$ at large $k R$. Its transformation \eqref{eq:S3toS2transform} yields $\Sinv(q)$ which is a good approximation of $\Sflat(q)$ if $R$ is not too small. The symbols $q$ and $k$ represent magnitudes of two- and three-dimensional wavevectors, respectively. Numbers in parentheses denote equation numbers of the relevant relationship.
}
\end{figure}

\section{\label{sec:relationship}Planar vs. Spherical Geometry}

We start with a planar, two-dimensional material that we assume to occupy the $xy$-plane in a Cartesian coordinate system. Its properties, in particular those that would be measured in a hypothetical scattering experiment, are described by the scalar field $\phi(x,y)$ or, equivalently, by its Fourier transform
\begin{equation}
\tilde{\phi}(\vek{q}) = \int \dif\vek{s}  \, \eexp{- i \vek{q}\vek{s}} \phi(\vek{s}) \label{eq:FourierTransform}
\end{equation}
where $\vek{s}=(x,y)$ and $\vek{q}=(q_x, q_y)$ denote points in two-dimensional position and reciprocal space, respectively.

The two-dimensional scattering intensity of this material is given by the structure factor
\begin{equation}
\Sflat (\vek{q}) \equiv \ffrac{1}{L^2} \mean{ \tilde{\phi}_{\vek{q}}  \tilde{\phi}_{-\vek{q} }}   \label{eq:S2d}
\end{equation}
which for isotropic materials depends only on the magnitude $q = \abs{\vek{q}}$ of the wavevector. Here $\mean{.}$ is the thermal equilibrium average, and we have included a normalization factor that removes the dependence on the membrane area $L^2$.

We now consider the embedding of the field $\phi(x,y)$ in three-dimensional Euclidean space,
\begin{equation}
\rho(\vek{r}) = \phi(x,y) \delta(z) .
\end{equation}
Here $\vek{r} = (x,y,z)$ is a point in three-dimensional space. This field has Fourier transform
\begin{eqnarray}
\tilde{\rho}(\vek{k})
& = & \int \dif \vek{r} \, \eexp{- i \vek{k}\vek{r}} \rho(\vek{r}) \\
& = & \tilde{\phi} (\vek{k}_\parallel)
\end{eqnarray}
where $\vek{k}_\parallel$ are the first two Cartesian components of the three-dimensional wavevector $\vek{k}$. The structure factor of the embedded field is
\begin{equation}
\Semb (\vek{k}) \equiv \ffrac{1}{L^2} \mean{\tilde{\rho}(\vek{k}) \tilde{\rho}(-\vek{k}) }
= \Sflat (\vek{k}_\parallel) .
\end{equation}
If the material is isotropic within the plane, then $\Semb (\vek{k})$ has \emph{cylindrical} symmetry in $\vek{k}$ space. We proceed by computing the \emph{spherical} average of this function over the polar angle $\theta$ and azimuthal angle $\varphi$:
\begin{eqnarray}
\Savg (k)
& \equiv & \ffrac{1}{4\pi} \int_0^\pi \sin \theta \, \dif \theta \int_0^{2\pi} \dif \varphi \, \Semb (\vek{k}) \\
& = & 
\ffrac{1}{k} \int_0^k \dif q \, \ffrac{q}{\sqrt{k^2-q^2}} \Sflat (q) .\label{eq:S2toS3transform}
\end{eqnarray}
This expression describes the spherically averaged scattering intensity of a planar material. It can also be interpreted as the scattering intensity that one would observe if one were to rotate the planar material across all possible orientations relative to an incoming beam, as illustrated in Figure~\ref{fig:sketch}.

The transformation \eqref{eq:S2toS3transform} from $\Sflat(q)$ to $\Savg(k)$  resembles the Abel transform as defined by Bracewell~\cite{Bracewell00}, albeit with different integration boundaries. Even though it is obtained by computing an orientational average it can be inverted. The loss of information that usually occurs when taking averages is circumvented by the embedding in a higher-dimensional space. 
We show in the appendix that the inverse of \eqref{eq:S2toS3transform} is
\begin{equation}
\Sflat(q) = \ffrac{2}{\pi} \int_0^q \dif k \,  \ffrac{\dif (k \Savg (k))}{\dif k} \ffrac{1}{\sqrt{q^2-k^2}} \label{eq:S3toS2transform} .
\end{equation}
This expression allows us to compute the in-plane, two-dimensional structure factor from the orientationally averaged, three-dimensional structure factor.

Note that \eqref{eq:S3toS2transform} is an improper integral, with a weak divergence of the integrand as $k \rightarrow q$. This makes this expression inconvenient to use if the integration has to be carried out numerically. We can avoid this difficulty by defining an auxiliary function
\begin{equation}
g (k)  = k \, \Savg(k)
\end{equation}
which, upon substitution into \eqref{eq:S3toS2transform} and partial integration, yields
\begin{equation}
\Sflat(q) = g'(q)- \ffrac{2}{\pi}   \int_0^q \dif k \,  \arctan \left( \ffrac{k}{\sqrt{q^2-k^2}} \right) g''(k) .
\label{eq:S3toS2transform2}
\end{equation}
In this formulation the divergence has been avoided at the cost of requiring information about the second derivative of $g (k)$.

Lastly we consider a two-dimensional material, again characterized by the scalar field $\phi$, that has the geometry of a spherical surface. In other words, its embedding in three-dimensional space is given by
\begin{equation}
\rho(\vek{r}) = \delta(r-R) \phi(\theta,\varphi) \label{eq:embedding}
\end{equation}
where $R$ is the radius of the sphere, $\delta$ is the Dirac delta function, and $(r,\theta,\varphi)$ are the spherical coordinates of the point $\vek{r}$. A scattering experiment measures this field's structure factor
\begin{equation}
\Ssphere (\vek{k}) \equiv \ffrac{1}{4 \pi R^2} \mean{\tilde{\rho}(\vek{k}) \tilde{\rho}(-\vek{k}) } .  \label{eq:Ssphere}
\end{equation}
As before we have included a normalization based on the surface area of the sphere, and we assume that this function is isotropic, i.e., independent of the direction of the wavevector $\vek{k}$.

Our goal is to establish a relationship between the two-dimensional structure factor of a planar material, $\Sflat (q)$, and the three-dimensional structure factor of the same material on the surface of a sphere, $\Ssphere (k)$. We already established the mathematical equivalency of the former to the orientationally averaged, planar structure factor $\Savg (k)$, expressed by \eqref{eq:S2toS3transform} and \eqref{eq:S3toS2transform}. The latter, in turn, is related to $\Ssphere (k)$: if the radius $R$ of the sphere is large compared to the length scale $1/k$ at which one probes the structure, one can think of the sphere as a collection of locally flat regions that exist at all possible orientations relative to an incoming beam. Based on this physical argument alone one expects that in the limit of $k R \rightarrow \infty$,
\begin{equation}
\Ssphere (k)  \rightarrow \Savg (k) \label{eq:SsphapproxSavg} .
\end{equation}
While we cannot provide a formal proof of this intuitive relationship for the general case, we will provide in the next section a demonstration of its validity for a specific choice of the material $\phi$.

We now assume this convergence to be smooth so that for sufficiently large $k R$ we can consider these two structure functions to be approximately equal (the validity of this assumption will be explored in the following section). Because the transformation \eqref{eq:S3toS2transform} converts $\Savg (k)$ into $\Sflat(q)$, it should do the same, at least approximately, when applied to $\Ssphere (k)$. We therefore define
\begin{equation}
\Sinv(q) =  \ffrac{2}{\pi} \int_0^q \dif k \,  \ffrac{\dif (k \Ssphere (k))}{\dif k} \ffrac{1}{\sqrt{q^2-k^2}}   \label{eq:S3toS2transform3}
\end{equation}
and use it as an estimate for the two-dimensional structure factor of the planar system,
\begin{equation}
\Sinv(q) \approx \Sflat(q) . \label{eq:invapproxflat}
\end{equation}
It is not possible to assess the accuracy of this approximation without making additional assumptions about the material under consideration. In the following section we will therefore evaluate the relationships derived so far in the context of a specific material that allows a complete analysis.

\section{\label{sec:gaussianmaterial}Application: A Gaussian material}

We consider a two-dimensional material described by its scattering length density $\phi(\vek{s})$ whose fluctuations are determined by the energy functional
\begin{equation}
E[\phi(\vek{s})] = \int \dif S \, \ffrac{\alpha}{2} \phi(\vek{s})^2 + \ffrac{\sigma}{2} \abs{\nabla \phi(\vek{s})}^2 + \ffrac{\kappa}{2} \left( \nabla^2 \phi (\vek{s})\right)^2 \label{eq:LGgeneral} ,
\end{equation}
where $\alpha$, $\sigma$ and $\kappa$ are parameters of the model, and the integral is taken over the entire area of the material. We consider separately the cases of planar and spherical materials. This model contains the quadratic terms of the well-known Landau-Ginzburg model, which has recently been used to study the phase behavior of multi-component lipid bilayers~\cite{Shlomovitz14, Sapp14, Luo18b} in addition to many other applications.

\subsection{Planar geometry}

If the surface is planar then it is convenient to expand the field $\phi(\vek{s})$ in a plane wave basis set, with expansion coefficients given by the Fourier transform \eqref{eq:FourierTransform}. In this representation the energy can be written as
\begin{equation}
E(\{\tilde{\phi}_{\vek{q}}\})  =  \ffrac{1}{2 L^2} \sum_{\vek{q}} \left( \alpha + \sigma q^2 + \kappa q^4 \right) \abs{\tilde{\phi}_{\vek{q}}}^2 \label{eq:GaussianEnergyinFourierSpace} 
\end{equation}
where we assumed the material to span a square patch of side length $L$, and the sum ranges over all wavevectors $\vek{q}$ consistent with that geometry. Note that the Fourier coefficients $\tilde{\phi}_{\vek{q}}$ act essentially like independent harmonic oscillators; the only coupling being the constraint $\tilde{\phi}_{\vek{q}} = \tilde{\phi}_{-\vek{q}}^*$ that ensures the field $\phi$ to be real-valued. The structure factor \eqref{eq:S2d} is therefore easily obtained from the equipartition theorem:
\begin{equation}
\Sflat (\vek{q}) =  \ffrac{\kT }{  \alpha + \sigma q^2 + \kappa q^4}  \label{eq:S2dGaussian} .
\end{equation}
Here $\kT$ is the thermal energy scale.

The structure factor must be positive for all values of the wavevector, which implies that both $\alpha$ and $\kappa$ must be positive. If $\sigma$ is also positive then $\Sflat$ is a monotonically decreasing function of $q$, which describes the fluctuations in an ordinary fluid. However, $\sigma$ can also be negative as long as 
\begin{equation}
\sigma > - 2 \sqrt{\alpha \kappa} . \label{eq:negativesigmabound}
\end{equation}
In this case the structure factor has a peak at wavevector 
\begin{equation}
\qc = \sqrt{-\sigma / 2\kappa} . \label{eq:qcflat}
\end{equation}
Such a peak is the hallmark characteristic of a microemulsion~\cite{Schick12,Shlomovitz13, Shlomovitz14}. The length $2\pi/\qc$ describes the scale of transient fluctuations and is therefore of fundamental interest.

\subsection{Spherical geometry}

If the material is shaped like the surface of a sphere then an expansion of the field $\phi$ in terms of spherical harmonic functions $Y_{l,m}$ is convenient:
\begin{equation}
\phi (\theta,\varphi) = \ffrac{1}{R^2} \sum_{l=0}^{\infty} \sum_{m=-l}^l u_{l,m} Y_{l,m} (\theta,\varphi) \label{eq:inverseSHTransform} .
\end{equation}
The complex coefficients $u_{l,m}$ can be obtained from the spherical harmonic transform
\begin{equation}
u_{l,m} = R^2  \int \dif \Omega \, Y_{l,m}^*  (\theta,\varphi) \phi (\theta,\varphi) ,
\end{equation}
where the integration is performed over all solid angles $\Omega$, and satisfy the condition $u_{l,-m} = (-1)^m u_{l,m}^*$. We have included a factor $R^2$ in the definition of the spherical harmonic coefficients $u_{l,m}$ to highlight their similarity to the Fourier coefficients $\tilde{\phi}_{\vek{q}}$ of the planar case.

Expressed in terms of $u_{l,m}$ the energy \eqref{eq:LGgeneral} becomes
\ifpreprint{
\begin{equation}
E(\{u_{l,m}\})  =  \ffrac{1}{2 R^2}  \sum_{l=0}^{\infty} \sum_{m=-l}^l  \bigg( \alpha + \ffrac{\sigma}{R^2} l (l+1)  
 + \ffrac{\kappa}{R^4}  \left( l (l+1)\right)^2 \bigg)  \abs{u_{l,m}}^2 \label{eq:GaussianEnergyonSphericalHarmonics}
\end{equation}
}
{
\begin{eqnarray}
E(\{u_{l,m}\}) & = &   \ffrac{1}{2 R^2}  \sum_{l=0}^{\infty} \sum_{m=-l}^l  \bigg( \alpha + \ffrac{\sigma}{R^2} l (l+1)  \label{eq:GaussianEnergyonSphericalHarmonics} \\*
&& \qquad \qquad {} + \ffrac{\kappa}{R^4}  \left( l (l+1)\right)^2 \bigg)  \abs{u_{l,m}}^2 \nonumber
\end{eqnarray}
}
and we can again use the equipartition theorem to determine the variance 
\begin{equation}
\mean{\abs{u_{l,m}}^2  } = \ffrac{\kT R^2}{\alpha + \ffrac{\sigma}{R^2} l (l+1) + \ffrac{\kappa}{R^4}  \left( l (l+1)\right)^2 } . \label{eq:ulmvariance}
\end{equation}

To compute the structure factor $\Ssphere (\vek{k})$ we use the embedding \eqref{eq:embedding} together with the expansion \eqref{eq:inverseSHTransform} to obtain the three-dimensional field $\rho(\vek{r})$. We calculate its Fourier transform by first expanding the plane wave basis set in spherical harmonics,
\begin{equation}
\eexp{-i\vek{k}\vek{r}}
=
4\pi \sum_{l=0}^\infty \sum_{m=-l}^l (-i)^{l} j_{l}(kr) Y_{l,m}(\hat{\vek{k}}) Y_{l,m}^* (\hat{\vek{r}}) ,
\end{equation}
and by using the orthonormality of the $Y_{l,m}$ to obtain
\begin{equation}
\tilde{\rho}(\vek{k})
=
4 \pi  \sum_{l=0}^{\infty} \sum_{m=-l}^l (-i)^l j_l(kR) Y_{l,m}(\hat{\vek{k}}) u_{l,m} .
\end{equation}
In these equations $j_l$ is the spherical Bessel function of the first kind, and $\hat{\vek{k}}$ is the unit vector in the direction of $\vek{k}$. This expression, together with \eqref{eq:ulmvariance}, can be used to compute the structure factor \eqref{eq:Ssphere},
\begin{equation}
\Ssphere(\vek{k}) =  \kT  \sum_{l=0}^\infty  \ffrac{(2l+1) j_l(kR)^2}{\alpha + \ffrac{\sigma}{R^2} l (l+1) + \ffrac{\kappa}{R^4}  \left( l (l+1)\right)^2 } , \label{eq:SsphereLG}
\end{equation}
where again we used the orthogonality of the spherical harmonic functions as well as Uns\"old's theorem, 
\begin{equation}
 \sum_{m=-l}^l  Y_{l,m}(\nvek{k}) Y_{l,m}(\nvek{k})^*
 =
 \ffrac{2l+1}{4\pi} .
 \end{equation}

A complete analysis of \eqref{eq:SsphereLG} is difficult due to the presence of the infinite series. However, we can obtain useful insight by considering the $k \rightarrow 0$ limit of $\Ssphere$ and its derivatives. We find that
\begin{eqnarray}
\Ssphere(k \rightarrow 0) & = & \kT  / \alpha ,\\
\ffrac{\dif \Ssphere } {\dif k} \bigg|_{k \rightarrow 0} & = & 0 ,\\
\ffrac{\dif^2 \Ssphere } {\dif k^2} \bigg|_{k \rightarrow 0} & = &
-   \ffrac{2 \kT R^2 }{3 \alpha} \left(  \ffrac{ \ffrac{2 \sigma}{R^2} + \ffrac{4 \kappa}{R^4}  }{\alpha + \ffrac{2 \sigma}{R^2} + \ffrac{4 \kappa}{R^4}}  \right)  . \label{eq:d2Sspheredk2}
\end{eqnarray}
If all parameters are positive then \eqref{eq:d2Sspheredk2} is negative, which together with the fact that $\Ssphere$ approaches zero at large wavevectors suggests that it is a monotonically decreasing function of $k$. We have already seen that $\sigma$ can be moderately negative as long as \eqref{eq:negativesigmabound} is satisfied. For a planar material this results in a peak in the structure factor at the non-zero wavevector $\qc$. In the spherical geometry we see that $\eqref{eq:d2Sspheredk2}$ is positive for $-2 \sqrt{\alpha \kappa} < \sigma < 0$ only if 
\begin{equation}
R > \sqrt{-2 \kappa/\sigma} \equiv \Rmin . \label{eq:Rmin}
\end{equation}
In this case the structure factor is positive and convex at $k = 0$, and converges to zero at large $k$. This implies the existence of a maximum at an intermediate wavevector, as is the case for a planar material. However, if $R < \Rmin$ then \eqref{eq:d2Sspheredk2} remains negative even for negative $\sigma$, which suggests that $\Ssphere$ remains a monotonic function without a peak. Because the presence of a peak in the structure factor is the identifying property of a microemulsion, we see that the assignment of thermodynamic phases can depend on the radius of the vesicle. This is one example of finite size effects that become apparent at small system sizes~\cite{Luo18b}.

This example illustrates that the geometry of a material (i.e., whether it is planar or spherical) can significantly change the scattering profile, and even result in qualitatively different interpretations (microemulsion or unstructured fluid) of its behavior. It follows that one should not directly compare  $\Sflat(q)$ and $\Ssphere(k)$. The transformation \eqref{eq:S3toS2transform3} serves as an approximate relationship that makes these structure factors comparable. To illustrate the accuracy of this transformation and to test the boundaries of its applicability we will now examine several examples of a material described by the energy \eqref{eq:LGgeneral}.

\subsection{Transformation between planar and spherical geometries}

Figure~\ref{fig:sofk} shows a comparison of the four different structure factors of interest: $\Sflat(q)$, the two-dimensional structure factor of a planar system, $\Ssphere(k)$, the three-dimensional structure factor of a spherical system, $\Sinv(q)$, the result of the approximate transformation of the spherical into a planar structure factor, and $\Savg(k)$, the structure factor of a planar system averaged over all possible orientations in three-dimensional space. Because the transformation is performed numerically we use the form \eqref{eq:S3toS2transform2} for its evaluation,
\begin{equation}
\Sinv(q) = g'(q)- \ffrac{2}{\pi}   \int_0^q \dif k \,  \arctan \left( \ffrac{k}{\sqrt{q^2-k^2}} \right) g''(k) .
\label{eq:S3toS2transform4}
\end{equation}
where $g(k) = k  \, \Ssphere(k)$.

Figure~\ref{fig:sofk}a illustrates that for non-negative values of the system parameters the structure factors of both the planar and the spherical system are monotonically decreasing. While those functions are qualitatively similar, they are not identical. The transformation \eqref{eq:S3toS2transform4} allows to compute the planar structure factor from the spherical one almost exactly ($\Sinv(q) \approx \Sflat(q)$), which demonstrates the utility of this approach. Here we chose a relatively large radius for the spherical system. As anticipated, the structure factor of the spherical system is essentially indistinguishable from that of the orientationally averaged planar system ($\Savg(k) \approx \Ssphere(k)$).

The effect of making $\sigma$ negative is shown Figure~\ref{fig:sofk}b. Both the planar and the spherical structure factor now exhibit a peak at non-zero wavevector, indicating that the material is now a microemulsion. Note that these peaks occur at different values of the wavevector. Inferring the length scale of the fluctuations present in this material from the inverse of the peak location in $\Ssphere(k)$ would yield a systematic underestimate. However, the transformation \eqref{eq:S3toS2transform4} allows to correct the effect of spherical geometry on the structure function, and to recover the intrinsic two-dimensional structure factor $\Sflat(q)$.

The accuracy of the transformation decreases when the radius of the spherical system becomes small, as illustrated in Figure~\ref{fig:sofk}c. Transforming $\Ssphere(k)$ into $\Sinv(q)$ no longer recovers $\Sflat(q)$ accurately, even though the location of the peak still shifts in the right direction, toward smaller wavevectors. Similarly one starts to see deviations between $\Ssphere(k)$ and the orientationally averaged structure factor of the planar system, $\Savg(k)$. 

Upon reducing the vesicle radius even further to $\Rmin$ the approximation \eqref{eq:invapproxflat} breaks down (Figure~\ref{fig:sofk}d). As expected for this radius, the structure factor of the spherical system no longer exhibits a peak at non-zero wavevector, and the transformation \eqref{eq:S3toS2transform4} does not recover it. Instead it generates a function that takes on negative values, which is not physical. For such small systems neither $\Sinv(q)$ nor $\Ssphere(k)$ are good approximations to $\Sflat(q)$.

\begin{figure}[tb]
\includegraphics[clip,trim=0in 0.72in 0in 0in]{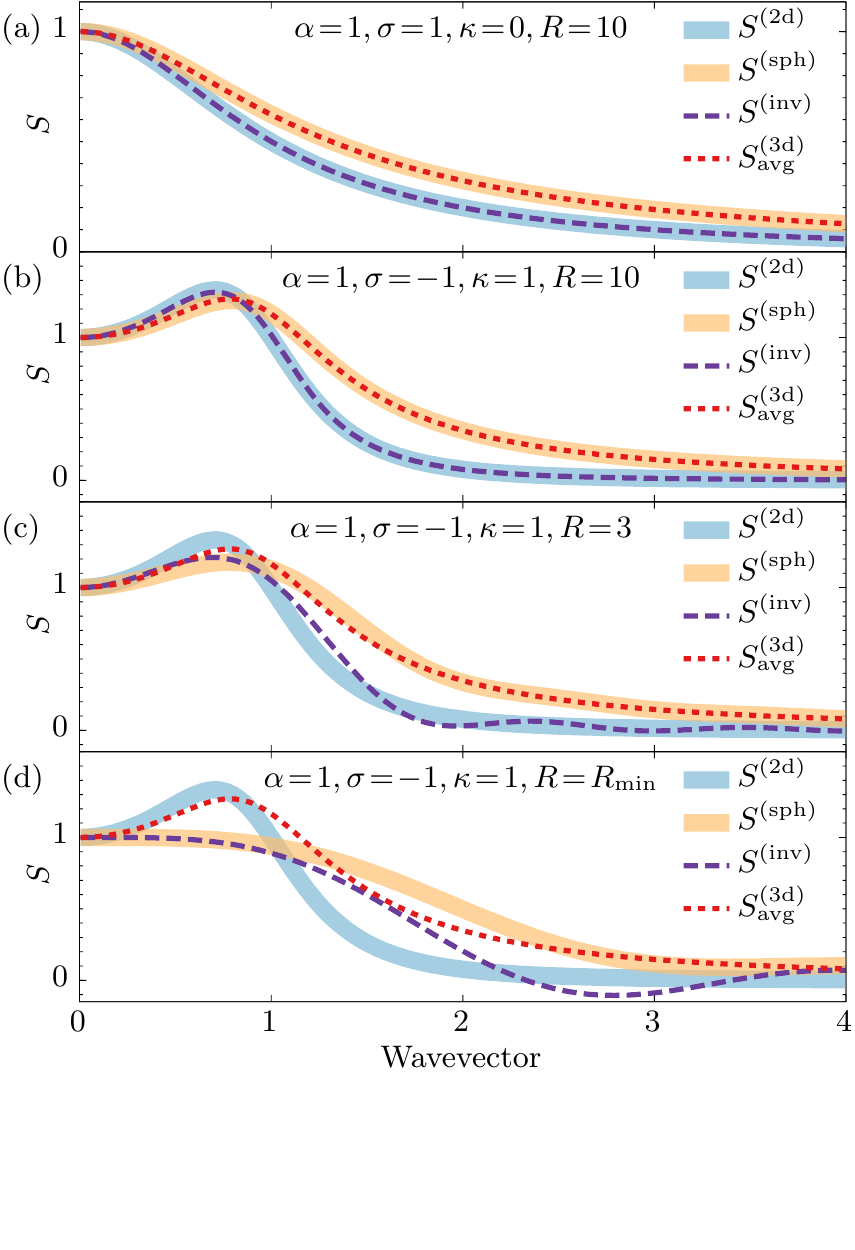}
\caption{\label{fig:sofk}Comparison of structure factors for different values of the dimensionless model parameters. See main text for a detailed discussion.  All calculations were performed at temperature $T\!=\!1$. The spherical structure factor $\Ssphere(k)$ was computed using \eqref{eq:SsphereLG} by including terms up to $l=1000$.}
\end{figure}

\section{\label{sec:discussion}Discussion}

The example showcased in the previous section illustrates both the benefits and the limitations of the proposed transformation of structure factors between spherical and planar systems, equation~\eqref{eq:S3toS2transform3}. If the radius of the sphere is large compared to the intrinsic length scale of the material then the transformation yields the planar structure factor nearly exactly. At intermediate radii the transformation is no longer exact, but at least qualitatively changes the structure factor in the right direction: $\Sinv$ is more similar to $\Sflat$ than $\Ssphere$ is. This is true generally for the overall shape of the scattering function as well as for important characteristic features, for example the location of a peak in the microemulsion regime.

At still smaller radii the transformation ceases to yield results that are similar to the planar structure factor, and the obtained $\Sinv(q)$ differs qualitatively from $\Sflat(q)$. Important features such as the existence of a peak at finite wavevectors are not preserved. In this regime the transformation~\eqref{eq:S3toS2transform3} should not be applied to estimate planar from spherical structure factors.
From~\eqref{eq:qcflat} and \eqref{eq:Rmin} we see that $\Rmin = 1 / \qc$, which has an appealing intuitive interpretation: in order to observe the length scale of the heterogeneity in a spherical system, the latter must be large enough to accommodate those fluctuations. In other words, the length of the sphere's great circle, $2 \pi R$, must exceed the correlation length $2\pi/\qc$ of the material.

In addition to the requirement of a spherical radius that exceeds the correlation length there are several other limitations that must be considered before using equation~\eqref{eq:S3toS2transform3} to compare planar and spherical structure factors. Those arise from the simplicity of the model proposed in this work. In particular, it treats the material as a two-dimensional surface, and ignores effects on the scattering intensity that originate from the non-zero thickness of the material. These effects become important for very small spheres when the thickness is a sizable fraction of the diameter. Another limitation is the model's restriction to perfectly planar and spherical geometries. Shape deformations, induced either by thermal fluctuations~\cite{Engelhardt85, Milner87} or by coupling between the composition field and the local geometry of the material, are not considered here. The latter are believed to play an important role in lipid bilayer mixtures that separate into domains with different spontaneous curvatures and/or bending rigidities~\cite{Baumgart03,Kawakatsu93,Taniguchi94,Taniguchi96,Shimobayashi16}.  Incorporating these effects into the model will further increase its usefulness in comparing the results of scattering experiments and/or calculations in planar and spherical geometries.

\section{\label{sec:conclusion}Conclusion}

We have developed an approximate relationship between planar and spherical structure factors of two-dimensional materials. We have demonstrated the accuracy and the limitations of this relationship by applying it to a Gaussian material that has previously been proposed to model multicomponent lipid bilayers and that supports both homogeneous and microemulsion phases. Our results allow a direct, model-free comparison of structure factors obtained in these different geometries.

\appendix*
\section{Derivation of Inverse Transform}

Equation \eqref{eq:S2toS3transform} defines $\Savg(k)$ in terms of $\Sflat(q)$. Here we show that the transformation \eqref{eq:S3toS2transform} is the inverse operation, i.e., it allows us to recover $\Sflat(q)$ from $\Savg(k)$.

We begin with the definition of $\Savg(k)$, equation \eqref{eq:S2toS3transform}. Using integration by parts, we find
\begin{equation}
k \Savg(k) =  k \Sflat(0) + \int_0^k \dif q \, \sqrt{k^2-q^2} \,\ffrac{\dif \Sflat(q)}{\dif q} .
\end{equation}
We now take the derivative with respect to $k$, keeping in mind that both the integrand and the integration boundary depend on that variable. However, the integrand is zero at $q=k$, and the Leibniz rule yields
\ifpreprint{
\begin{equation}
\ffrac{\dif}{\dif k} \left(k \Savg(k) \right) =
\Sflat(0) + \int_0^k \dif q \, \ffrac{\partial}{\partial k} \left( \sqrt{k^2-q^2}  \right) \, \ffrac{\dif S^{\text{(2d)}}(q)}{\dif q} . 
 \end{equation}
}
{
\begin{eqnarray}
\ffrac{\dif}{\dif k} \left(k \Savg(k) \right) &=&
\Sflat(0) \\
&& \hspace{-4em} {} +  \int_0^k \dif q \, \ffrac{\partial}{\partial k} \left( \sqrt{k^2-q^2}  \right) \, \ffrac{\dif S^{\text{(2d)}}(q)}{\dif q} . \nonumber
 \end{eqnarray}
}
The derivative in the integrand is easily evaluated to yield $k/\sqrt{k^2 - q^2} $.
We now multiply both sides of this equation by $1/\sqrt{p^2 - k^2}$ and integrate from 0 to $p$, where $p$ is an arbitrary non-negative number:
\ifpreprint{
\begin{eqnarray}
\int_0^p \dif k \ffrac{\dif}{\dif k} \left( k \Savg (k) \right)  \ffrac{1}{\sqrt{p^2-k^2}}
&=&
 S^{\text{(2d)}}(0) \int_0^p \dif k   \ffrac{1}{\sqrt{p^2-k^2}} \\*
 && {} + \int_0^p \dif k  \ffrac{1}{\sqrt{p^2-k^2}} \int_0^k \dif q \ffrac{k}{\sqrt{k^2-q^2}}  \ffrac{\dif \Sflat(q)}{\dif q}  . \nonumber
\end{eqnarray}
}
{
\begin{widetext}
\begin{equation}
\int_0^p \dif k \ffrac{\dif}{\dif k} \left( k \Savg (k) \right)  \ffrac{1}{\sqrt{p^2-k^2}}
=
 S^{\text{(2d)}}(0) \int_0^p \dif k   \ffrac{1}{\sqrt{p^2-k^2}}
+ \int_0^p \dif k  \ffrac{1}{\sqrt{p^2-k^2}} \int_0^k \dif q \ffrac{k}{\sqrt{k^2-q^2}}  \ffrac{\dif \Sflat(q)}{\dif q}  .
\end{equation}
\end{widetext}
}
The value of the first integral on the right-hand side is $\pi/2$. We now rewrite the double integral $\int_0^p \dif k \, \int_0^k \dif q$ as $\int_0^p \dif q \, \int_q^p \dif k$. This allows us to isolate the $k$-integral, which again yields $\pi/2$. The above expression is therefore equal to
\begin{equation}
\ffrac{\pi}{2} \left( \Sflat(0) +   \int_0^p \dif q \ffrac{\dif \Sflat(q)}{\dif q} \right)
=
\ffrac{\pi}{2}  \Sflat(p) .
\end{equation}
After multiplying both sides by $2/\pi$ and relabeling $p \rightarrow q$ we arrive at \eqref{eq:S3toS2transform}.




\bibliography{paper}

\begin{thebibliography}{39}%
\makeatletter
\providecommand \@ifxundefined [1]{%
 \@ifx{#1\undefined}
}%
\providecommand \@ifnum [1]{%
 \ifnum #1\expandafter \@firstoftwo
 \else \expandafter \@secondoftwo
 \fi
}%
\providecommand \@ifx [1]{%
 \ifx #1\expandafter \@firstoftwo
 \else \expandafter \@secondoftwo
 \fi
}%
\providecommand \natexlab [1]{#1}%
\providecommand \enquote  [1]{``#1''}%
\providecommand \bibnamefont  [1]{#1}%
\providecommand \bibfnamefont [1]{#1}%
\providecommand \citenamefont [1]{#1}%
\providecommand \href@noop [0]{\@secondoftwo}%
\providecommand \href [0]{\begingroup \@sanitize@url \@href}%
\providecommand \@href[1]{\@@startlink{#1}\@@href}%
\providecommand \@@href[1]{\endgroup#1\@@endlink}%
\providecommand \@sanitize@url [0]{\catcode `\\12\catcode `\$12\catcode
  `\&12\catcode `\#12\catcode `\^12\catcode `\_12\catcode `\%12\relax}%
\providecommand \@@startlink[1]{}%
\providecommand \@@endlink[0]{}%
\providecommand \url  [0]{\begingroup\@sanitize@url \@url }%
\providecommand \@url [1]{\endgroup\@href {#1}{\urlprefix }}%
\providecommand \urlprefix  [0]{URL }%
\providecommand \Eprint [0]{\href }%
\providecommand \doibase [0]{http://dx.doi.org/}%
\providecommand \selectlanguage [0]{\@gobble}%
\providecommand \bibinfo  [0]{\@secondoftwo}%
\providecommand \bibfield  [0]{\@secondoftwo}%
\providecommand \translation [1]{[#1]}%
\providecommand \BibitemOpen [0]{}%
\providecommand \bibitemStop [0]{}%
\providecommand \bibitemNoStop [0]{.\EOS\space}%
\providecommand \EOS [0]{\spacefactor3000\relax}%
\providecommand \BibitemShut  [1]{\csname bibitem#1\endcsname}%
\let\auto@bib@innerbib\@empty
\bibitem [{\citenamefont {Veatch}\ and\ \citenamefont
  {Keller}(2003)}]{Veatch03}%
  \BibitemOpen
  \bibfield  {author} {\bibinfo {author} {\bibfnamefont {S.~L.}\ \bibnamefont
  {Veatch}}\ and\ \bibinfo {author} {\bibfnamefont {S.~L.}\ \bibnamefont
  {Keller}},\ }\bibfield  {title} {\enquote {\bibinfo {title} {Separation of
  liquid phases in giant vesicles of ternary mixtures of phospholipids and
  cholesterol},}\ }\href {\doibase 10.1016/S0006-3495(03)74726-2} {\bibfield
  {journal} {\bibinfo  {journal} {Biophys. J.}\ }\textbf {\bibinfo {volume}
  {85}},\ \bibinfo {pages} {3074--3083} (\bibinfo {year} {2003})}\BibitemShut
  {NoStop}%
\bibitem [{\citenamefont {Baumgart}, \citenamefont {Hess},\ and\ \citenamefont
  {Webb}(2003)}]{Baumgart03}%
  \BibitemOpen
  \bibfield  {author} {\bibinfo {author} {\bibfnamefont {T.}~\bibnamefont
  {Baumgart}}, \bibinfo {author} {\bibfnamefont {S.~T.}\ \bibnamefont {Hess}},
  \ and\ \bibinfo {author} {\bibfnamefont {W.~W.}\ \bibnamefont {Webb}},\
  }\bibfield  {title} {\enquote {\bibinfo {title} {Imaging coexisting fluid
  domains in biomembrane models coupling curvature and line tension},}\ }\href
  {\doibase 10.1038/nature02013} {\bibfield  {journal} {\bibinfo  {journal}
  {Nature}\ }\textbf {\bibinfo {volume} {425}},\ \bibinfo {pages} {821--824}
  (\bibinfo {year} {2003})}\BibitemShut {NoStop}%
\bibitem [{\citenamefont {Scherfeld}, \citenamefont {Kahya},\ and\
  \citenamefont {Schwille}(2003)}]{Scherfeld03}%
  \BibitemOpen
  \bibfield  {author} {\bibinfo {author} {\bibfnamefont {D.}~\bibnamefont
  {Scherfeld}}, \bibinfo {author} {\bibfnamefont {N.}~\bibnamefont {Kahya}}, \
  and\ \bibinfo {author} {\bibfnamefont {P.}~\bibnamefont {Schwille}},\
  }\bibfield  {title} {\enquote {\bibinfo {title} {Lipid dynamics and domain
  formation in model membranes composed of ternary mixtures of unsaturated and
  saturated phosphatidylcholines and cholesterol},}\ }\href {\doibase
  10.1016/S0006-3495(03)74791-2} {\bibfield  {journal} {\bibinfo  {journal}
  {Biophys. J.}\ }\textbf {\bibinfo {volume} {85}},\ \bibinfo {pages}
  {3758--3768} (\bibinfo {year} {2003})}\BibitemShut {NoStop}%
\bibitem [{\citenamefont {Veatch}\ and\ \citenamefont
  {Keller}(2005)}]{Veatch05}%
  \BibitemOpen
  \bibfield  {author} {\bibinfo {author} {\bibfnamefont {S.~L.}\ \bibnamefont
  {Veatch}}\ and\ \bibinfo {author} {\bibfnamefont {S.~L.}\ \bibnamefont
  {Keller}},\ }\bibfield  {title} {\enquote {\bibinfo {title} {Seeing spots:
  Complex phase behavior in simple membranes},}\ }\href {\doibase
  10.1016/j.bbamcr.2005.06.010} {\bibfield  {journal} {\bibinfo  {journal}
  {Biochim. Biophys. Acta -- Molecular Cell Research}\ }\textbf {\bibinfo
  {volume} {1746}},\ \bibinfo {pages} {172--185} (\bibinfo {year}
  {2005})}\BibitemShut {NoStop}%
\bibitem [{\citenamefont {Marsh}(2009)}]{Marsh09}%
  \BibitemOpen
  \bibfield  {author} {\bibinfo {author} {\bibfnamefont {D.}~\bibnamefont
  {Marsh}},\ }\bibfield  {title} {\enquote {\bibinfo {title}
  {Cholesterol-induced fluid membrane domains: A compendium of lipid-raft
  ternary phase diagrams},}\ }\href {\doibase 10.1016/j.bbamem.2009.08.004}
  {\bibfield  {journal} {\bibinfo  {journal} {Biochim. Biophys. Acta --
  Biomembranes}\ }\textbf {\bibinfo {volume} {1788}},\ \bibinfo {pages}
  {2114--2123} (\bibinfo {year} {2009})}\BibitemShut {NoStop}%
\bibitem [{\citenamefont {Kumar}, \citenamefont {Gompper},\ and\ \citenamefont
  {Lipowsky}(1999)}]{Kumar99}%
  \BibitemOpen
  \bibfield  {author} {\bibinfo {author} {\bibfnamefont {P.~B.~S.}\
  \bibnamefont {Kumar}}, \bibinfo {author} {\bibfnamefont {G.}~\bibnamefont
  {Gompper}}, \ and\ \bibinfo {author} {\bibfnamefont {R.}~\bibnamefont
  {Lipowsky}},\ }\bibfield  {title} {\enquote {\bibinfo {title} {Modulated
  phases in multicomponent fluid membranes},}\ }\href {\doibase
  10.1103/PhysRevE.60.4610} {\bibfield  {journal} {\bibinfo  {journal} {Phys.
  Rev. E}\ }\textbf {\bibinfo {volume} {60}},\ \bibinfo {pages} {4610--4618}
  (\bibinfo {year} {1999})}\BibitemShut {NoStop}%
\bibitem [{\citenamefont {Schick}(2012)}]{Schick12}%
  \BibitemOpen
  \bibfield  {author} {\bibinfo {author} {\bibfnamefont {M.}~\bibnamefont
  {Schick}},\ }\bibfield  {title} {\enquote {\bibinfo {title} {Membrane
  heterogeneity: Manifestation of a curvature-induced microemulsion},}\ }\href
  {\doibase 10.1103/PhysRevE.85.031902} {\bibfield  {journal} {\bibinfo
  {journal} {Phys. Rev. E}\ }\textbf {\bibinfo {volume} {85}},\ \bibinfo
  {pages} {031902} (\bibinfo {year} {2012})}\BibitemShut {NoStop}%
\bibitem [{\citenamefont {Shlomovitz}, \citenamefont {Maibaum},\ and\
  \citenamefont {Schick}(2014)}]{Shlomovitz14}%
  \BibitemOpen
  \bibfield  {author} {\bibinfo {author} {\bibfnamefont {R.}~\bibnamefont
  {Shlomovitz}}, \bibinfo {author} {\bibfnamefont {L.}~\bibnamefont {Maibaum}},
  \ and\ \bibinfo {author} {\bibfnamefont {M.}~\bibnamefont {Schick}},\
  }\bibfield  {title} {\enquote {\bibinfo {title} {Macroscopic phase
  separation, modulated phases, and microemulsions: A unified picture of
  rafts},}\ }\href {\doibase 10.1016/j.bpj.2014.03.017} {\bibfield  {journal}
  {\bibinfo  {journal} {Biophys. J.}\ }\textbf {\bibinfo {volume} {106}},\
  \bibinfo {pages} {1979--1985} (\bibinfo {year} {2014})}\BibitemShut {NoStop}%
\bibitem [{\citenamefont {Shimobayashi}, \citenamefont {Ichikawa},\ and\
  \citenamefont {Taniguchi}(2016)}]{Shimobayashi16}%
  \BibitemOpen
  \bibfield  {author} {\bibinfo {author} {\bibfnamefont {S.~F.}\ \bibnamefont
  {Shimobayashi}}, \bibinfo {author} {\bibfnamefont {M.}~\bibnamefont
  {Ichikawa}}, \ and\ \bibinfo {author} {\bibfnamefont {T.}~\bibnamefont
  {Taniguchi}},\ }\bibfield  {title} {\enquote {\bibinfo {title} {Direct
  observations of transition dynamics from macro- to micro-phase separation in
  asymmetric lipid bilayers induced by externally added glycolipids},}\ }\href
  {\doibase 10.1209/0295-5075/113/56005} {\bibfield  {journal} {\bibinfo
  {journal} {Europhys. Lett.}\ }\textbf {\bibinfo {volume} {113}},\ \bibinfo
  {pages} {56005} (\bibinfo {year} {2016})}\BibitemShut {NoStop}%
\bibitem [{\citenamefont {Feigenson}(2009)}]{Feigenson09}%
  \BibitemOpen
  \bibfield  {author} {\bibinfo {author} {\bibfnamefont {G.~W.}\ \bibnamefont
  {Feigenson}},\ }\bibfield  {title} {\enquote {\bibinfo {title} {Phase
  diagrams and lipid domains in multicomponent lipid bilayer mixtures},}\
  }\href {\doibase 10.1016/j.bbamem.2008.08.014} {\bibfield  {journal}
  {\bibinfo  {journal} {Biochim. Biophys. Acta -- Biomembranes}\ }\textbf
  {\bibinfo {volume} {1788}},\ \bibinfo {pages} {47--52} (\bibinfo {year}
  {2009})}\BibitemShut {NoStop}%
\bibitem [{\citenamefont {Konyakhina}\ \emph {et~al.}(2011)\citenamefont
  {Konyakhina}, \citenamefont {Goh}, \citenamefont {Amazon}, \citenamefont
  {Heberle}, \citenamefont {Wu},\ and\ \citenamefont
  {Feigenson}}]{Konyakhina11}%
  \BibitemOpen
  \bibfield  {author} {\bibinfo {author} {\bibfnamefont {T.~M.}\ \bibnamefont
  {Konyakhina}}, \bibinfo {author} {\bibfnamefont {S.~L.}\ \bibnamefont {Goh}},
  \bibinfo {author} {\bibfnamefont {J.}~\bibnamefont {Amazon}}, \bibinfo
  {author} {\bibfnamefont {F.~A.}\ \bibnamefont {Heberle}}, \bibinfo {author}
  {\bibfnamefont {J.}~\bibnamefont {Wu}}, \ and\ \bibinfo {author}
  {\bibfnamefont {G.~W.}\ \bibnamefont {Feigenson}},\ }\bibfield  {title}
  {\enquote {\bibinfo {title} {Control of a nanoscopic-to-macroscopic
  transition: {M}odulated phases in four-component {DSPC}/{DOPC}/{POPC}/{C}hol
  giant unilamellar vesicles},}\ }\href {\doibase 10.1016/j.bpj.2011.06.019}
  {\bibfield  {journal} {\bibinfo  {journal} {Biophys. J.}\ }\textbf {\bibinfo
  {volume} {101}},\ \bibinfo {pages} {L8--L10} (\bibinfo {year}
  {2011})}\BibitemShut {NoStop}%
\bibitem [{\citenamefont {Rheinst{\"a}dter}\ and\ \citenamefont
  {Mouritsen}(2013)}]{Rheinstadter13}%
  \BibitemOpen
  \bibfield  {author} {\bibinfo {author} {\bibfnamefont {M.~C.}\ \bibnamefont
  {Rheinst{\"a}dter}}\ and\ \bibinfo {author} {\bibfnamefont {O.~G.}\
  \bibnamefont {Mouritsen}},\ }\bibfield  {title} {\enquote {\bibinfo {title}
  {Small-scale structure in fluid cholesterol--lipid bilayers},}\ }\href
  {\doibase 10.1016/j.cocis.2013.07.001} {\bibfield  {journal} {\bibinfo
  {journal} {Curr. Op. in Coll. \& Interf. Sci.}\ }\textbf {\bibinfo {volume}
  {18}},\ \bibinfo {pages} {440--447} (\bibinfo {year} {2013})}\BibitemShut
  {NoStop}%
\bibitem [{\citenamefont {Armstrong}\ \emph {et~al.}(2013)\citenamefont
  {Armstrong}, \citenamefont {Marquardt}, \citenamefont {Dies}, \citenamefont
  {Kucerka}, \citenamefont {Yamani}, \citenamefont {Harroun}, \citenamefont
  {Katsaras}, \citenamefont {Shi},\ and\ \citenamefont
  {Rheinst{\"a}dter}}]{Armstrong13}%
  \BibitemOpen
  \bibfield  {author} {\bibinfo {author} {\bibfnamefont {C.~L.}\ \bibnamefont
  {Armstrong}}, \bibinfo {author} {\bibfnamefont {D.}~\bibnamefont
  {Marquardt}}, \bibinfo {author} {\bibfnamefont {H.}~\bibnamefont {Dies}},
  \bibinfo {author} {\bibfnamefont {N.}~\bibnamefont {Kucerka}}, \bibinfo
  {author} {\bibfnamefont {Z.}~\bibnamefont {Yamani}}, \bibinfo {author}
  {\bibfnamefont {T.~A.}\ \bibnamefont {Harroun}}, \bibinfo {author}
  {\bibfnamefont {J.}~\bibnamefont {Katsaras}}, \bibinfo {author}
  {\bibfnamefont {A.-C.}\ \bibnamefont {Shi}}, \ and\ \bibinfo {author}
  {\bibfnamefont {M.~C.}\ \bibnamefont {Rheinst{\"a}dter}},\ }\bibfield
  {title} {\enquote {\bibinfo {title} {The observation of highly ordered
  domains in membranes with cholesterol},}\ }\href {\doibase
  10.1371/journal.pone.0066162} {\bibfield  {journal} {\bibinfo  {journal}
  {PLoS One}\ }\textbf {\bibinfo {volume} {8}},\ \bibinfo {pages} {e66162}
  (\bibinfo {year} {2013})}\BibitemShut {NoStop}%
\bibitem [{\citenamefont {Heberle}\ \emph {et~al.}(2013)\citenamefont
  {Heberle}, \citenamefont {Petruzielo}, \citenamefont {Pan}, \citenamefont
  {Drazba}, \citenamefont {Kucerka}, \citenamefont {Standaert}, \citenamefont
  {Feigenson},\ and\ \citenamefont {Katsaras}}]{Heberle13b}%
  \BibitemOpen
  \bibfield  {author} {\bibinfo {author} {\bibfnamefont {F.~A.}\ \bibnamefont
  {Heberle}}, \bibinfo {author} {\bibfnamefont {R.~S.}\ \bibnamefont
  {Petruzielo}}, \bibinfo {author} {\bibfnamefont {J.}~\bibnamefont {Pan}},
  \bibinfo {author} {\bibfnamefont {P.}~\bibnamefont {Drazba}}, \bibinfo
  {author} {\bibfnamefont {N.}~\bibnamefont {Kucerka}}, \bibinfo {author}
  {\bibfnamefont {R.~F.}\ \bibnamefont {Standaert}}, \bibinfo {author}
  {\bibfnamefont {G.~W.}\ \bibnamefont {Feigenson}}, \ and\ \bibinfo {author}
  {\bibfnamefont {J.}~\bibnamefont {Katsaras}},\ }\bibfield  {title} {\enquote
  {\bibinfo {title} {Bilayer thickness mismatch controls domain size in model
  membranes},}\ }\href {\doibase 10.1021/ja3113615} {\bibfield  {journal}
  {\bibinfo  {journal} {J. Am. Chem. Soc.}\ }\textbf {\bibinfo {volume}
  {135}},\ \bibinfo {pages} {6853--6859} (\bibinfo {year} {2013})}\BibitemShut
  {NoStop}%
\bibitem [{\citenamefont {Toppozini}\ \emph {et~al.}(2014)\citenamefont
  {Toppozini}, \citenamefont {Meinhardt}, \citenamefont {Armstrong},
  \citenamefont {Yamani}, \citenamefont {Kucerka}, \citenamefont {Schmid},\
  and\ \citenamefont {Rheinst{\"a}dter}}]{Toppozini14}%
  \BibitemOpen
  \bibfield  {author} {\bibinfo {author} {\bibfnamefont {L.}~\bibnamefont
  {Toppozini}}, \bibinfo {author} {\bibfnamefont {S.}~\bibnamefont
  {Meinhardt}}, \bibinfo {author} {\bibfnamefont {C.~L.}\ \bibnamefont
  {Armstrong}}, \bibinfo {author} {\bibfnamefont {Z.}~\bibnamefont {Yamani}},
  \bibinfo {author} {\bibfnamefont {N.}~\bibnamefont {Kucerka}}, \bibinfo
  {author} {\bibfnamefont {F.}~\bibnamefont {Schmid}}, \ and\ \bibinfo {author}
  {\bibfnamefont {M.~C.}\ \bibnamefont {Rheinst{\"a}dter}},\ }\bibfield
  {title} {\enquote {\bibinfo {title} {Structure of cholesterol in lipid
  rafts},}\ }\href {\doibase 10.1103/PhysRevLett.113.228101} {\bibfield
  {journal} {\bibinfo  {journal} {Phys. Rev. Lett.}\ }\textbf {\bibinfo
  {volume} {113}},\ \bibinfo {pages} {228101} (\bibinfo {year}
  {2014})}\BibitemShut {NoStop}%
\bibitem [{\citenamefont {Nickels}\ \emph {et~al.}(2015)\citenamefont
  {Nickels}, \citenamefont {Cheng}, \citenamefont {Mostofian}, \citenamefont
  {Stanley}, \citenamefont {Lindner}, \citenamefont {Heberle}, \citenamefont
  {Perticaroli}, \citenamefont {Feygenson}, \citenamefont {Egami},
  \citenamefont {Standaert}, \citenamefont {Smith}, \citenamefont {Myles},
  \citenamefont {Ohl},\ and\ \citenamefont {Katsaras}}]{Nickels15}%
  \BibitemOpen
  \bibfield  {author} {\bibinfo {author} {\bibfnamefont {J.~D.}\ \bibnamefont
  {Nickels}}, \bibinfo {author} {\bibfnamefont {X.}~\bibnamefont {Cheng}},
  \bibinfo {author} {\bibfnamefont {B.}~\bibnamefont {Mostofian}}, \bibinfo
  {author} {\bibfnamefont {C.}~\bibnamefont {Stanley}}, \bibinfo {author}
  {\bibfnamefont {B.}~\bibnamefont {Lindner}}, \bibinfo {author} {\bibfnamefont
  {F.~A.}\ \bibnamefont {Heberle}}, \bibinfo {author} {\bibfnamefont
  {S.}~\bibnamefont {Perticaroli}}, \bibinfo {author} {\bibfnamefont
  {M.}~\bibnamefont {Feygenson}}, \bibinfo {author} {\bibfnamefont
  {T.}~\bibnamefont {Egami}}, \bibinfo {author} {\bibfnamefont {R.~F.}\
  \bibnamefont {Standaert}}, \bibinfo {author} {\bibfnamefont {J.~C.}\
  \bibnamefont {Smith}}, \bibinfo {author} {\bibfnamefont {D.~A.~A.}\
  \bibnamefont {Myles}}, \bibinfo {author} {\bibfnamefont {M.}~\bibnamefont
  {Ohl}}, \ and\ \bibinfo {author} {\bibfnamefont {J.}~\bibnamefont
  {Katsaras}},\ }\bibfield  {title} {\enquote {\bibinfo {title} {Mechanical
  properties of nanoscopic lipid domains},}\ }\href {\doibase
  10.1021/jacs.5b08894} {\bibfield  {journal} {\bibinfo  {journal} {J. Am.
  Chem. Soc.}\ }\textbf {\bibinfo {volume} {137}},\ \bibinfo {pages}
  {15772--15780} (\bibinfo {year} {2015})}\BibitemShut {NoStop}%
\bibitem [{\citenamefont {Enoki}, \citenamefont {Heberle},\ and\ \citenamefont
  {Feigenson}(2018)}]{Enoki18}%
  \BibitemOpen
  \bibfield  {author} {\bibinfo {author} {\bibfnamefont {T.~A.}\ \bibnamefont
  {Enoki}}, \bibinfo {author} {\bibfnamefont {F.~A.}\ \bibnamefont {Heberle}},
  \ and\ \bibinfo {author} {\bibfnamefont {G.~W.}\ \bibnamefont {Feigenson}},\
  }\bibfield  {title} {\enquote {\bibinfo {title} {{FRET} detects the size of
  nanodomains for coexisting liquid-disordered and liquid-ordered phases},}\
  }\href {\doibase 10.1016/j.bpj.2018.03.014} {\bibfield  {journal} {\bibinfo
  {journal} {Biophys. J.}\ }\textbf {\bibinfo {volume} {114}},\ \bibinfo
  {pages} {1921--1935} (\bibinfo {year} {2018})}\BibitemShut {NoStop}%
\bibitem [{\citenamefont {Hansen}\ and\ \citenamefont
  {McDonald}(2013)}]{HansenMcDonald13}%
  \BibitemOpen
  \bibfield  {author} {\bibinfo {author} {\bibfnamefont {J.-P.}\ \bibnamefont
  {Hansen}}\ and\ \bibinfo {author} {\bibfnamefont {I.~R.}\ \bibnamefont
  {McDonald}},\ }\href@noop {} {\emph {\bibinfo {title} {Theory of Simple
  Liquids}}},\ \bibinfo {edition} {4th}\ ed.\ (\bibinfo  {publisher} {Academic
  Press},\ \bibinfo {address} {Oxford},\ \bibinfo {year} {2013})\BibitemShut
  {NoStop}%
\bibitem [{\citenamefont {Pabst}\ \emph {et~al.}(2010)\citenamefont {Pabst},
  \citenamefont {Kucerka}, \citenamefont {Nieh}, \citenamefont
  {Rheinst{\"a}dter},\ and\ \citenamefont {Katsaras}}]{Pabst10}%
  \BibitemOpen
  \bibfield  {author} {\bibinfo {author} {\bibfnamefont {G.}~\bibnamefont
  {Pabst}}, \bibinfo {author} {\bibfnamefont {N.}~\bibnamefont {Kucerka}},
  \bibinfo {author} {\bibfnamefont {M.-P.}\ \bibnamefont {Nieh}}, \bibinfo
  {author} {\bibfnamefont {M.~C.}\ \bibnamefont {Rheinst{\"a}dter}}, \ and\
  \bibinfo {author} {\bibfnamefont {J.}~\bibnamefont {Katsaras}},\ }\bibfield
  {title} {\enquote {\bibinfo {title} {Applications of neutron and {X}-ray
  scattering to the study of biologically relevant model membranes},}\ }\href
  {\doibase 10.1016/j.chemphyslip.2010.03.010} {\bibfield  {journal} {\bibinfo
  {journal} {Chem. Phys. Lipids}\ }\textbf {\bibinfo {volume} {163}},\ \bibinfo
  {pages} {460--479} (\bibinfo {year} {2010})}\BibitemShut {NoStop}%
\bibitem [{\citenamefont {Marquardt}\ \emph {et~al.}(2015)\citenamefont
  {Marquardt}, \citenamefont {Heberle}, \citenamefont {Nickels}, \citenamefont
  {Pabst},\ and\ \citenamefont {Katsaras}}]{Marquardt15}%
  \BibitemOpen
  \bibfield  {author} {\bibinfo {author} {\bibfnamefont {D.}~\bibnamefont
  {Marquardt}}, \bibinfo {author} {\bibfnamefont {F.~A.}\ \bibnamefont
  {Heberle}}, \bibinfo {author} {\bibfnamefont {J.~D.}\ \bibnamefont
  {Nickels}}, \bibinfo {author} {\bibfnamefont {G.}~\bibnamefont {Pabst}}, \
  and\ \bibinfo {author} {\bibfnamefont {J.}~\bibnamefont {Katsaras}},\
  }\bibfield  {title} {\enquote {\bibinfo {title} {On scattered waves and lipid
  domains: detecting membrane rafts with x-rays and neutrons},}\ }\href
  {\doibase 10.1039/c5sm01807b} {\bibfield  {journal} {\bibinfo  {journal}
  {Soft Matter}\ }\textbf {\bibinfo {volume} {11}},\ \bibinfo {pages}
  {9055--9072} (\bibinfo {year} {2015})}\BibitemShut {NoStop}%
\bibitem [{\citenamefont {Hirose}, \citenamefont {Komura},\ and\ \citenamefont
  {Andelman}(2012)}]{Hirose12b}%
  \BibitemOpen
  \bibfield  {author} {\bibinfo {author} {\bibfnamefont {Y.}~\bibnamefont
  {Hirose}}, \bibinfo {author} {\bibfnamefont {S.}~\bibnamefont {Komura}}, \
  and\ \bibinfo {author} {\bibfnamefont {D.}~\bibnamefont {Andelman}},\
  }\bibfield  {title} {\enquote {\bibinfo {title} {Concentration fluctuations
  and phase transitions in coupled modulated bilayers},}\ }\href {\doibase
  10.1103/PhysRevE.86.021916} {\bibfield  {journal} {\bibinfo  {journal} {Phys.
  Rev. E}\ }\textbf {\bibinfo {volume} {86}},\ \bibinfo {pages} {021916}
  (\bibinfo {year} {2012})}\BibitemShut {NoStop}%
\bibitem [{\citenamefont {Shlomovitz}\ and\ \citenamefont
  {Schick}(2013)}]{Shlomovitz13}%
  \BibitemOpen
  \bibfield  {author} {\bibinfo {author} {\bibfnamefont {R.}~\bibnamefont
  {Shlomovitz}}\ and\ \bibinfo {author} {\bibfnamefont {M.}~\bibnamefont
  {Schick}},\ }\bibfield  {title} {\enquote {\bibinfo {title} {Model of a raft
  in both leaves of an asymmetric lipid bilayer},}\ }\href {\doibase
  10.1016/j.bpj.2013.06.053} {\bibfield  {journal} {\bibinfo  {journal}
  {Biophys. J.}\ }\textbf {\bibinfo {volume} {105}},\ \bibinfo {pages}
  {1406--1413} (\bibinfo {year} {2013})}\BibitemShut {NoStop}%
\bibitem [{\citenamefont {Palmieri}\ and\ \citenamefont
  {Safran}(2013)}]{Palmieri13}%
  \BibitemOpen
  \bibfield  {author} {\bibinfo {author} {\bibfnamefont {B.}~\bibnamefont
  {Palmieri}}\ and\ \bibinfo {author} {\bibfnamefont {S.~A.}\ \bibnamefont
  {Safran}},\ }\bibfield  {title} {\enquote {\bibinfo {title} {Hybrid lipids
  increase the probability of fluctuating nanodomains in mixed membranes},}\
  }\href {\doibase 10.1021/la4006168} {\bibfield  {journal} {\bibinfo
  {journal} {Langmuir}\ }\textbf {\bibinfo {volume} {29}},\ \bibinfo {pages}
  {5246--5261} (\bibinfo {year} {2013})}\BibitemShut {NoStop}%
\bibitem [{\citenamefont {Rosetti}\ and\ \citenamefont
  {Pastorino}(2012)}]{Rosetti12}%
  \BibitemOpen
  \bibfield  {author} {\bibinfo {author} {\bibfnamefont {C.}~\bibnamefont
  {Rosetti}}\ and\ \bibinfo {author} {\bibfnamefont {C.}~\bibnamefont
  {Pastorino}},\ }\bibfield  {title} {\enquote {\bibinfo {title} {Comparison of
  ternary bilayer mixtures with asymmetric or symmetric unsaturated
  phosphatidylcholine lipids by coarse grained molecular dynamics
  simulations},}\ }\href {\doibase 10.1021/jp212406u} {\bibfield  {journal}
  {\bibinfo  {journal} {J. Phys. Chem. B}\ }\textbf {\bibinfo {volume} {116}},\
  \bibinfo {pages} {3525--3537} (\bibinfo {year} {2012})}\BibitemShut {NoStop}%
\bibitem [{\citenamefont {Ackerman}\ and\ \citenamefont
  {Feigenson}(2015)}]{Ackerman15}%
  \BibitemOpen
  \bibfield  {author} {\bibinfo {author} {\bibfnamefont {D.~G.}\ \bibnamefont
  {Ackerman}}\ and\ \bibinfo {author} {\bibfnamefont {G.~W.}\ \bibnamefont
  {Feigenson}},\ }\bibfield  {title} {\enquote {\bibinfo {title} {Multiscale
  modeling of four-component lipid mixtures: Domain composition, size,
  alignment, and properties of the phase interface},}\ }\href {\doibase
  10.1021/jp511083z} {\bibfield  {journal} {\bibinfo  {journal} {J. Phys. Chem.
  B}\ }\textbf {\bibinfo {volume} {119}},\ \bibinfo {pages} {4240--4250}
  (\bibinfo {year} {2015})}\BibitemShut {NoStop}%
\bibitem [{\citenamefont {Baoukina}, \citenamefont {Rozmanov},\ and\
  \citenamefont {Tieleman}(2017)}]{Baoukina17}%
  \BibitemOpen
  \bibfield  {author} {\bibinfo {author} {\bibfnamefont {S.}~\bibnamefont
  {Baoukina}}, \bibinfo {author} {\bibfnamefont {D.}~\bibnamefont {Rozmanov}},
  \ and\ \bibinfo {author} {\bibfnamefont {D.~P.}\ \bibnamefont {Tieleman}},\
  }\bibfield  {title} {\enquote {\bibinfo {title} {Composition fluctuations in
  lipid bilayers},}\ }\href {\doibase 10.1016/j.bpj.2017.10.009} {\bibfield
  {journal} {\bibinfo  {journal} {Biophys. J.}\ }\textbf {\bibinfo {volume}
  {113}},\ \bibinfo {pages} {2750--2761} (\bibinfo {year} {2017})}\BibitemShut
  {NoStop}%
\bibitem [{\citenamefont {He}\ and\ \citenamefont {Maibaum}(2018)}]{He18}%
  \BibitemOpen
  \bibfield  {author} {\bibinfo {author} {\bibfnamefont {S.}~\bibnamefont
  {He}}\ and\ \bibinfo {author} {\bibfnamefont {L.}~\bibnamefont {Maibaum}},\
  }\bibfield  {title} {\enquote {\bibinfo {title} {Identifying the onset of
  phase separation in quaternary lipid bilayer systems from coarse-grained
  simulations},}\ }\href {\doibase 10.1021/acs.jpcb.8b00364} {\bibfield
  {journal} {\bibinfo  {journal} {J. Phys. Chem. B}\ }\textbf {\bibinfo
  {volume} {122}},\ \bibinfo {pages} {3961--3973} (\bibinfo {year}
  {2018})}\BibitemShut {NoStop}%
\bibitem [{\citenamefont {Pencer}\ \emph {et~al.}(2005)\citenamefont {Pencer},
  \citenamefont {Mills}, \citenamefont {Anghel}, \citenamefont {Krueger},
  \citenamefont {Epand},\ and\ \citenamefont {Katsaras}}]{Pencer05b}%
  \BibitemOpen
  \bibfield  {author} {\bibinfo {author} {\bibfnamefont {J.}~\bibnamefont
  {Pencer}}, \bibinfo {author} {\bibfnamefont {T.}~\bibnamefont {Mills}},
  \bibinfo {author} {\bibfnamefont {V.}~\bibnamefont {Anghel}}, \bibinfo
  {author} {\bibfnamefont {S.}~\bibnamefont {Krueger}}, \bibinfo {author}
  {\bibfnamefont {R.~M.}\ \bibnamefont {Epand}}, \ and\ \bibinfo {author}
  {\bibfnamefont {J.}~\bibnamefont {Katsaras}},\ }\bibfield  {title} {\enquote
  {\bibinfo {title} {Detection of submicron-sized raft-like domains in
  membranes by small-angle neutron scattering},}\ }\href {\doibase
  10.1140/epje/e2005-00046-5} {\bibfield  {journal} {\bibinfo  {journal} {Eur.
  Phys. J. E}\ }\textbf {\bibinfo {volume} {18}},\ \bibinfo {pages} {447--458}
  (\bibinfo {year} {2005})}\BibitemShut {NoStop}%
\bibitem [{\citenamefont {Jiang}\ and\ \citenamefont {Powers}(2008)}]{Jiang08}%
  \BibitemOpen
  \bibfield  {author} {\bibinfo {author} {\bibfnamefont {H.}~\bibnamefont
  {Jiang}}\ and\ \bibinfo {author} {\bibfnamefont {T.~R.}\ \bibnamefont
  {Powers}},\ }\bibfield  {title} {\enquote {\bibinfo {title} {Curvature-driven
  lipid sorting in a membrane tubule},}\ }\href {\doibase
  10.1103/PhysRevLett.101.018103} {\bibfield  {journal} {\bibinfo  {journal}
  {Phys. Rev. Lett.}\ }\textbf {\bibinfo {volume} {101}},\ \bibinfo {pages}
  {018103} (\bibinfo {year} {2008})}\BibitemShut {NoStop}%
\bibitem [{\citenamefont {Sakuma}\ \emph {et~al.}(2011)\citenamefont {Sakuma},
  \citenamefont {Urakami}, \citenamefont {Taniguchi},\ and\ \citenamefont
  {Imai}}]{Sakuma11}%
  \BibitemOpen
  \bibfield  {author} {\bibinfo {author} {\bibfnamefont {Y.}~\bibnamefont
  {Sakuma}}, \bibinfo {author} {\bibfnamefont {N.}~\bibnamefont {Urakami}},
  \bibinfo {author} {\bibfnamefont {T.}~\bibnamefont {Taniguchi}}, \ and\
  \bibinfo {author} {\bibfnamefont {M.}~\bibnamefont {Imai}},\ }\bibfield
  {title} {\enquote {\bibinfo {title} {Asymmetric distribution of cone-shaped
  lipids in a highly curved bilayer revealed by a small angle neutron
  scattering technique},}\ }\href {\doibase 10.1088/0953-8984/23/28/284104}
  {\bibfield  {journal} {\bibinfo  {journal} {J. Phys. Cond. Mat.}\ }\textbf
  {\bibinfo {volume} {23}},\ \bibinfo {pages} {284104} (\bibinfo {year}
  {2011})}\BibitemShut {NoStop}%
\bibitem [{\citenamefont {Callan-Jones}, \citenamefont {Sorre},\ and\
  \citenamefont {Bassereau}(2011)}]{CallanJones11}%
  \BibitemOpen
  \bibfield  {author} {\bibinfo {author} {\bibfnamefont {A.}~\bibnamefont
  {Callan-Jones}}, \bibinfo {author} {\bibfnamefont {B.}~\bibnamefont {Sorre}},
  \ and\ \bibinfo {author} {\bibfnamefont {P.}~\bibnamefont {Bassereau}},\
  }\bibfield  {title} {\enquote {\bibinfo {title} {Curvature-driven lipid
  sorting in biomembranes},}\ }\href {\doibase 10.1101/cshperspect.a004648}
  {\bibfield  {journal} {\bibinfo  {journal} {Cold Spring Harbor Perspectives
  in Biology}\ }\textbf {\bibinfo {volume} {3}},\ \bibinfo {pages} {a004648}
  (\bibinfo {year} {2011})}\BibitemShut {NoStop}%
\bibitem [{\citenamefont {Bracewell}(2000)}]{Bracewell00}%
  \BibitemOpen
  \bibfield  {author} {\bibinfo {author} {\bibfnamefont {R.~N.}\ \bibnamefont
  {Bracewell}},\ }\href@noop {} {\emph {\bibinfo {title} {The Fourier Transform
  and Its Applications}}},\ \bibinfo {edition} {3rd}\ ed.\ (\bibinfo
  {publisher} {McGraw-Hill},\ \bibinfo {year} {2000})\BibitemShut {NoStop}%
\bibitem [{\citenamefont {Sapp}, \citenamefont {Shlomovitz},\ and\
  \citenamefont {Maibaum}(2014)}]{Sapp14}%
  \BibitemOpen
  \bibfield  {author} {\bibinfo {author} {\bibfnamefont {K.}~\bibnamefont
  {Sapp}}, \bibinfo {author} {\bibfnamefont {R.}~\bibnamefont {Shlomovitz}}, \
  and\ \bibinfo {author} {\bibfnamefont {L.}~\bibnamefont {Maibaum}},\
  }\bibfield  {title} {\enquote {\bibinfo {title} {Seeing the forest in lieu of
  the trees: Continuum simulations of cell membranes at large length scales},}\
  }in\ \href {\doibase 10.1016/B978-0-444-63378-1.00003-3} {\emph {\bibinfo
  {booktitle} {Annual Reports in Computational Chemistry}}},\ Vol.~\bibinfo
  {volume} {10},\ \bibinfo {editor} {edited by\ \bibinfo {editor}
  {\bibfnamefont {R.~A.}\ \bibnamefont {Wheeler}}}\ (\bibinfo  {publisher}
  {Elsevier},\ \bibinfo {address} {Amsterdam},\ \bibinfo {year} {2014})\ pp.\
  \bibinfo {pages} {47--76}\BibitemShut {NoStop}%
\bibitem [{\citenamefont {Luo}\ and\ \citenamefont {Maibaum}()}]{Luo18b}%
  \BibitemOpen
  \bibfield  {author} {\bibinfo {author} {\bibfnamefont {Y.}~\bibnamefont
  {Luo}}\ and\ \bibinfo {author} {\bibfnamefont {L.}~\bibnamefont {Maibaum}},\
  }\href@noop {} {}\bibinfo {howpublished} {in preparation}\BibitemShut
  {NoStop}%
\bibitem [{\citenamefont {Engelhardt}, \citenamefont {Duwe},\ and\
  \citenamefont {Sackmann}(1985)}]{Engelhardt85}%
  \BibitemOpen
  \bibfield  {author} {\bibinfo {author} {\bibfnamefont {H.}~\bibnamefont
  {Engelhardt}}, \bibinfo {author} {\bibfnamefont {H.~P.}\ \bibnamefont
  {Duwe}}, \ and\ \bibinfo {author} {\bibfnamefont {E.}~\bibnamefont
  {Sackmann}},\ }\bibfield  {title} {\enquote {\bibinfo {title} {Bilayer
  bending elasticity measured by fourier analysis of thermally excited surface
  undulations of flaccid vesicles},}\ }\href {\doibase
  10.1051/jphyslet:01985004608039500} {\bibfield  {journal} {\bibinfo
  {journal} {J. Physique Lett.}\ }\textbf {\bibinfo {volume} {46}},\ \bibinfo
  {pages} {395--400} (\bibinfo {year} {1985})}\BibitemShut {NoStop}%
\bibitem [{\citenamefont {Milner}\ and\ \citenamefont
  {Safran}(1987)}]{Milner87}%
  \BibitemOpen
  \bibfield  {author} {\bibinfo {author} {\bibfnamefont {S.~T.}\ \bibnamefont
  {Milner}}\ and\ \bibinfo {author} {\bibfnamefont {S.~A.}\ \bibnamefont
  {Safran}},\ }\bibfield  {title} {\enquote {\bibinfo {title} {Dynamical
  fluctuations of droplet microemulsions and vesicles},}\ }\href {\doibase
  10.1103/PhysRevA.36.4371} {\bibfield  {journal} {\bibinfo  {journal} {Phys.
  Rev. A}\ }\textbf {\bibinfo {volume} {36}},\ \bibinfo {pages} {4371--4379}
  (\bibinfo {year} {1987})}\BibitemShut {NoStop}%
\bibitem [{\citenamefont {Kawakatsu}\ \emph {et~al.}(1993)\citenamefont
  {Kawakatsu}, \citenamefont {Andelman}, \citenamefont {Kawasaki},\ and\
  \citenamefont {Taniguchi}}]{Kawakatsu93}%
  \BibitemOpen
  \bibfield  {author} {\bibinfo {author} {\bibfnamefont {T.}~\bibnamefont
  {Kawakatsu}}, \bibinfo {author} {\bibfnamefont {D.}~\bibnamefont {Andelman}},
  \bibinfo {author} {\bibfnamefont {K.}~\bibnamefont {Kawasaki}}, \ and\
  \bibinfo {author} {\bibfnamefont {T.}~\bibnamefont {Taniguchi}},\ }\bibfield
  {title} {\enquote {\bibinfo {title} {Phase transitions and shapes of two
  component membranes and vesicles {I}: strong segregation limit},}\ }\href
  {\doibase 10.1051/jp2:1993177} {\bibfield  {journal} {\bibinfo  {journal} {J.
  Phys. II (France)}\ }\textbf {\bibinfo {volume} {3}},\ \bibinfo {pages}
  {971--997} (\bibinfo {year} {1993})}\BibitemShut {NoStop}%
\bibitem [{\citenamefont {Taniguchi}\ \emph {et~al.}(1994)\citenamefont
  {Taniguchi}, \citenamefont {Kawasaki}, \citenamefont {Andelman},\ and\
  \citenamefont {Kawakatsu}}]{Taniguchi94}%
  \BibitemOpen
  \bibfield  {author} {\bibinfo {author} {\bibfnamefont {T.}~\bibnamefont
  {Taniguchi}}, \bibinfo {author} {\bibfnamefont {K.}~\bibnamefont {Kawasaki}},
  \bibinfo {author} {\bibfnamefont {D.}~\bibnamefont {Andelman}}, \ and\
  \bibinfo {author} {\bibfnamefont {T.}~\bibnamefont {Kawakatsu}},\ }\bibfield
  {title} {\enquote {\bibinfo {title} {Phase transitions and shapes of two
  component membranes and vesicles {II}: weak segregation limit},}\ }\href
  {\doibase 10.1051/jp2:1994203} {\bibfield  {journal} {\bibinfo  {journal} {J.
  Phys. II (France)}\ }\textbf {\bibinfo {volume} {4}},\ \bibinfo {pages}
  {1333--1362} (\bibinfo {year} {1994})}\BibitemShut {NoStop}%
\bibitem [{\citenamefont {Taniguchi}(1996)}]{Taniguchi96}%
  \BibitemOpen
  \bibfield  {author} {\bibinfo {author} {\bibfnamefont {T.}~\bibnamefont
  {Taniguchi}},\ }\bibfield  {title} {\enquote {\bibinfo {title} {Shape
  deformation and phase separation dynamics of two-component vesicles},}\
  }\href {\doibase 10.1103/PhysRevLett.76.4444} {\bibfield  {journal} {\bibinfo
   {journal} {Phys. Rev. Lett.}\ }\textbf {\bibinfo {volume} {76}},\ \bibinfo
  {pages} {4444--4447} (\bibinfo {year} {1996})}\BibitemShut {NoStop}%
\end{thebibliography}%

\end{document}